\documentclass[twocolumn, showpacs,preprintnumbers,%
amsmath,amssymb,prc]{revtex4}
\begin{document}
\preprint{GVKA-1(2008)}
\title{$\omega$ meson production in   $ pp $ collisions with a polarized beam}
\author{J Balasubramanyam}
 \affiliation{Department of Physics, Bangalore University, Bangalore-560056, India \\
 K. S. Institute of Technology, Bangalore-560062, India}
\author{Venkataraya}
 \affiliation{ Vijaya College, Bangalore-560011, India}
\author{G Ramachandran}
 \affiliation{GVK Academy,  Jayanagar,  Bangalore - 560082, India}

\date{\today}

\begin{abstract}
Model independent formulae are derived for the beam analyzing power $A_y$ and beam to 
meson spin transfers in $pp \to pp \omega$ taking into consideration all the six
threshold partial wave  amplitudes covering the $Ss, \,Sp$ and $Ps$ channels. Attention is 
also focussed on the empirical determination of the lowest three partial wave amplitudes
$f_1, f_2 ,f_3$ without any discrete ambiguities.
\end{abstract}
\pacs{13.25.-k, 13.60.Le, 13.75.-n, 13.88.+e, 24.70.+s, 25.40.ve}
\maketitle

Meson production in $N-N$ collisions has excited considerable interest \cite{mac}, ever since
the measurements \cite{meyo} in the early 1990's revealed that the total cross-section for $pp \to \ pp\pi^0$
exceeded the then available theoretical estimates \cite{kol} by more than a factor of 5. Moreover, a large 
momentum transfer is involved when an additional particle is produced in the final state, which 
implies that the features of the $N-N$ interaction is probed at very short distances. These have 
been estimated \cite{nak} to be of the order of $0.53fm, \, 0.21fm$ and $0.18fm$ for the production of 
$\pi, \, \omega $ and $\varphi$ respectively. The experimental studies in the case of pion production 
have reached a high degree of sophistication \cite{mey,von},  where the three body final state is completely
identified kinematically and spin observables are measured employing a polarized beam on a polarized 
target. The J\"ulich meson exchange model  \cite{han},  which yielded theoretical predictions 
closer to data than most other models,  has been more successful in the case of charged 
pion production \cite{von}  than with neutral pions \cite{mey}.  A recent analysis \cite{dep}  of
$\vec{p}\vec{p} \to pp \pi^0$  measurements \cite{mey}, following a model-independent irreducible tensor approach
\cite{gr1},  showed that the J\"ulich model deviates from the empirically extracted estimates quite
significantly for the $ ^3P_1 \; \to \;  ^3P_0p$ and to a lesser extent for the $^3F_3 \; \to \;  ^3P_2p$ 
transitions;  this analysis has also emphasized the importance of $\bigtriangleup$ 
contribution as the model calculations has also been carried out with and without 
taking into consideration the $\bigtriangleup$  contribution
to emphasize its importance in the model calculation. 
 In contrast, the production of isoscalar mesons $\omega$ and $\varphi$ involves 
 only excited nucleon states \cite{barn}.  Moreover, 
the Okubo-Zweig-Iizuka (OZI) rule \cite{oku}  suppresses $\varphi$  production relative to 
$\omega$ production. In view of the dramatic violation \cite{ams} 
of this rule observed in $\bar{p}p$
collisions, the ratio $R_{\phi/\omega}$ was measured~\cite{bal} and it was
found to be an order of magnitude larger, after correction for the available
phase space, than the theoretical estimate $R_{OZI} = 4.2 \times 10^{-3}$
~\cite{lip}. The latest experimental estimate~\cite{hart} is $R_{\phi/\omega} 
\approx 8 \times R_{OZI} $. The total cross-section for $pp \to pp\omega$ was measured 
\cite{hib} at 5 excess energies $\epsilon$ in the range $3.8 MeV$ to $30 MeV$ in c.m. 
The threshold energy dependence up to $\epsilon = 320 MeV$ has been studied using 
several models \cite{fal} theoretically. A strong anisotropic angular distribution was reported 
\cite{abd}  at $\epsilon = 173 MeV$ from an experimental study at the time-of-flight spectrometer
TOF of COoler SYnchrotron COSY \cite{kth} at Julich and the onset of higher partial waves 
was seen at a much lower energy in the more recent measurements \cite{bar1} at 
the COSY-ANKE facility \cite{bar2} and    \cite{mar}
at two values of $\epsilon$ higher than \cite{bar1}. Quark model calculations \cite{kon}
have also predicted anisotropy in the angular distribution.  A set of six partial wave amplitudes 
have been identified  \cite{gr3} to study the reaction at threshold and near threshold 
energies covering $Ss, \, Sp$ and $Ps$ channels. Taking into consideration only the $Ss$ and 
$Sp$ amplitudes,  the then existing data \cite{bal,abd,mar} on the 
 differential cross-section was  analyzed \cite{gr2},  where  it was also shown 
 that the empirical estimates of the three amplitudes could be obtained from experimental 
 measurements of the differential cross-section,  $\omega$  meson polarization and the 
 analyzing power in a polarized beam and  polarized  target experiment, for which
 a  proposal had already  been made \cite{rat}. 
Very recently the beam analyzing power $A_y$ was measured \cite{abl} for the first
time. A program to measure the beam to meson spin transfer is underway
\cite{bri}, using the $3\pi$ decay mode of $\omega$. In this context, it has   recently been
shown \cite{gr4} that the $3\pi$ decay mode of 
$\omega$ can be utilized to determine the tensor polarization of $\omega$,
but not its vector polarization.

The purpose of this communication is  to  extend the  
model independent theoretical approach \cite{gr3,gr2} to examine  (i) the analyzing power 
\cite{abl}  with  a  polarized beam and (ii) the beam to meson spin transfer \cite{bri}, 
taking into consideration all the six $Ss, \, Sp, \, Ps$ threshold partial wave  amplitudes.  
We also focus attention on the empirical determination of the lowest 
three amplitudes, free from discrete ambiguities, from such measurements \cite{abl,bri},
together with measurements employing an unpolarized beam.

The notation $(^{2s+1}L_j)_i \, \to  \, (^{2s+1}L_j)_f l $ used in \cite{mey}, 
to designate the partial wave amplitudes in 
the context of $pp \to pp\pi^o$ , is by itself inadequate to describe 
completely the partial wave amplitudes for $pp \to pp\omega$, since $\omega$
has spin 1 in contrast to the spin zero of the pion. Therefore, one has to 
either employ the notations  introduced earlier in \cite{gr3} or generalize the notations 
used in \cite{mey} to $(^{2s+1}L_j)_i \to (^{2s+1}L_j)_f \, ^3l_{j_{\omega}} $, 
together with the understanding that the vector addition of $j_{\omega}$ and $j_f$ 
yields $j_i =j$ in order to conserve the total angular momentum $j$ in the reaction. To 
facilitate comparison of the two different notations, we may now change the symbols 
$ l_i, l_f, L, S$ of \cite{gr3} to  $L_i, L_f, \mathcal{L}, \mathcal{S} $ 
respectively and note  that  the orbital angular momenta and spins have been added in \cite{gr3} in a
$L-S$ coupling scheme, in contrast to the generalization to \cite{mey} suggested above which
corresponds to $j-j$ coupling. We may, therefore, express the matrix elements 
$R_{\mathcal{L} \mathcal{S}} = \langle((l_{\omega}L_f)\mathcal{L}(1s_f)\mathcal{S})j\|T \| (L_is_i)j \rangle$
in terms of $M_{j_\omega j_f} =  \langle ((l_{\omega}1) j_{\omega}(L_fs_f)j_f)j \|T \|(L_is_i)j\rangle$
through
\begin{eqnarray}
 R_{\mathcal{L} \mathcal{S}} &=&
  [\mathcal{L}][\mathcal{S}]  \sum_{ j_{\omega}  j_f}  [j_{\omega}][j_f]
\left\{ \begin{array}{ccc}
              l_{\omega} & 1 & j_{\omega} \\
			  L_f & s_f & j_f \\
			  \mathcal{L} &  \mathcal{S} & j  
              \end{array} \right\}  M_{j_\omega  j_f},
\label{eq:comp}
\end{eqnarray}
and enumerate the lowest six threshold partial wave amplitudes covering the $Ss, \,Sp, \, Ps$ 
channels in the two schemes as $R_1,\ldots,R_6$ and $M_1,\ldots,M_6$ respectively. Using 
Eq.\eqref{eq:comp}, we have
\begin{eqnarray}R_k &=& M_k\;;\; k = 1,2,3 \; ; \; R_4=-M_4, \nonumber \\
R_5 &=& {1 \over 2}(M_5+\sqrt{3}M_6)\;;\; 
R_6= {1 \over 2}(\sqrt{3}M_5-M_6) 
\label{eq:gma}.
\end{eqnarray}

The $R_k$ as well as the $M_k$ are functions of c.m. energy $E$ at which the reaction 
takes place and the invariant mass $W$ of  the two protons system or 
equivalently energy $E_\omega$ of the $\omega$ meson produced in the final state.
The six $T^j_{\alpha \beta}$ in  Table 1 of \cite{gr3}, which were enumerated there as 
 $T_1,\ldots,T_6$ are proportional to  $R_1,\ldots,R_6$ as given by Eq.(5) of \cite{gr3}. The 
 $T_k$, when multiplied by a factor 
\begin{equation}
F = \left[{ WE_\omega (E -E_\omega)q p_f \over 4(2\pi)^5 p_i}\right]^{1/2}
\end{equation}
depending purely on the kinematical variables, lead to 
the partial wave amplitudes $f_k, \;\;\; ( k = 1,\ldots,6)$ employed in \cite{gr2},
 which are thus given explicited by
\begin{eqnarray}
\label{eq:fjb}
f_k = F\,T_k &=&   4  \left( \frac{2\pi WE_{\omega}(E-E_{\omega} )q p_f }{3   p_i} \right)^{1/2} 
\nonumber \\ && \times
(-1)^{\mathcal{L}+L_i+s_i-j} [j]^2[\mathcal{S}] [s_f]^{-1} R_k,   
\end{eqnarray}
where $E_{\omega}$ and ${\boldsymbol q}$ denote the  energy and momentum of the 
$\omega$ meson in the c.m frame,  while ${\boldsymbol p}_i$ and 
${\boldsymbol p}_f $ denote respectively the initial and final relative momenta between the two
protons in their respective c.m. frames such that $(q, \, \theta , \, \varphi) , \; (p_i , \theta_i , \varphi_i)
, \; (p_f, \theta_f , \varphi_f)$ denote the polar co-ordinates of   ${\boldsymbol q} , \,{\boldsymbol p}_i,
\,{\boldsymbol p}_f $ respectively. The $E_\omega, \, q, \, p_i \, $ and $p_f$ are known,  
if $E$ and $W$ are given. Defining 
\begin{equation}
f ' = f_2 +{1 \over \sqrt{10}}f_3  \;\;\; ; \;\;\; f'' = f_2 -{2 \over \sqrt{10}}f_3,
\end{equation} 
the unpolarized differential cross-section for $pp \to pp\omega$ 
may be expressed following \cite{gr2}, as 
\begin{eqnarray}
\frac{d\sigma_o}{dW d\Omega} &=&  {1 \over 4}\int d\Omega_f \,
Tr({\mathcal M}{\mathcal M}^\dagger)   \nonumber  \\ &=& 
{1 \over 192 \pi^2  } [\alpha_1 + 3 ( |f'|^2\,cos^2\theta + |f''|^2\,sin^2\theta)] 
\nonumber  \\ &=&   {1 \over 192 \pi^2  } 
[\alpha_0 + 0.9 \, \alpha_2 \,  cos^2\theta]
 \label{updcs},
\end{eqnarray}
where ${\mathcal M} = F \,{\boldsymbol T} $ denotes the reaction matrix, 
${\boldsymbol T} $ being the on-energy-shell transition matrix given by Eq.(3) of \cite{gr3} 
and ${\mathcal M}^\dagger$ denotes the  hermitian conjugate of ${\mathcal M}$. 
The coefficients $\alpha_1$ and  $\alpha_2 $  on the right hand side of Eq.\eqref{updcs}
are given by
\begin{eqnarray}
\alpha_1 &=& |f_1|^2+9|f_4|^2 +
{9\over 5}|f_5|^2+{27 \over 25 }|f_6|^2,
 \\
\alpha_0 &=& \alpha_1 + 3|f'|^2 \\
\alpha_2  &=& |f_3|^2-2 \sqrt{10}\,\Re( f_2f_3^*),
\end{eqnarray}
which reduce to those given by Eqs.(10) and (11) of \cite{gr2}, if we set $f_4 =f_5 = f_6 =0$. The
differential cross-section given by Eq.\eqref{updcs} may be multiplied by $(W/E)$ to yield the 
differential cross-section $( d\sigma_o / dE_\omega d\Omega)$.

If ${\boldsymbol P}$ denotes the polarization of the proton beam the spin density matrix 
$\rho^i$ characterizing the initial state may be written as 
\begin{equation}
\rho^i = \frac{1}{4}(1+ {\boldsymbol \sigma}_1 \cdot {\boldsymbol P} ),
\end{equation}
while the  density matrix $\rho^f $ for the final state is defined in terms of its elements
\begin{equation}
\rho^f_{\chi_f , \chi_f '}  =  \langle s_f m_f ; m|{\mathcal M} \rho^i {\mathcal M}^\dagger
|s_f'm_f' ; m'\rangle ,
\end{equation}
where $\chi_f \equiv (s_f , m_f,  m )$. The differential cross-section for 
$p(\vec{p},\omega)pp$ is given by 
\begin{equation}
\label{dcs}
\frac{d\sigma}{dW d\Omega} = \int d\Omega_f Tr\rho^f = \frac{d\sigma_o}{dW d\Omega}[1+{\boldsymbol P}
\cdot {\boldsymbol A} ],
\end{equation}
where the  analyzing power 
\begin{equation}
\label{ap}
{\boldsymbol A} ={ 1 \over 32 \sqrt{6} \;\pi^2}  \, \beta_1 ( \hat{\boldsymbol q} \times 
  \hat{\boldsymbol p}_i)  \; \; ; \; \;  \beta_1 =  \Im( f_1^* f'),
\end{equation}
 is transverse to the reaction plane and hence has a
single component  
\begin{equation}
A_y = -{ 1 \over 32 \sqrt{6} \;\pi^2}\Im( f_1^* f')\, sin\theta,
\end{equation}
in the Madison frame \cite{sat}. It may be noted that 
the z-axis of the right handed frame, refered  to above, is along the beam,  while ${\boldsymbol q}$
lies in the reaction plane i.e., z-x plane so that the azimuthal angle $\varphi$ of 
${\boldsymbol q}$ is zero. If ${\boldsymbol P}$ is transverse and if the x-axis is chosen along ${\boldsymbol P}$
so that $\varphi$ is not necessarily zero, the term ${\boldsymbol P} \cdot {\boldsymbol A}$
in Eq.\eqref{dcs} is given by 
\begin{equation}
{\boldsymbol P}\cdot {\boldsymbol A} = { P \over 32 \sqrt{6} \; \pi^2}\Im( f_1^* f') \, sin\theta\,sin\varphi.
\end{equation}
where $P = |{\boldsymbol P}|$.

The density matrix $\rho^\omega$ characterizing the spin state of the 
$\omega$ produced with c.m. energy $E_\omega$, 
when the beam is polarized,  is defined in terms of its elements by 
\begin{eqnarray}
\rho^\omega_{m, m '}  &=&  \sum_{s_f m_f} \int d\Omega_f  \; \rho^f_{\chi_f , \chi_f '} \\
&=& \frac{Tr\rho^\omega}{3}\sum_{k=0}^2(-1)^q C(1k1;m'-qm)[k] \, t^k_q,
\end{eqnarray}
so that the Fano statistical tensors $t^k_q, k=1,2$ (which define respectively the vector and tensor 
polarizations of the $\omega$ meson) are given, with respect to the Madison frame \cite{sat}, by
\begin{eqnarray}
\label{t10}
C \,t^1_0 &=&   \;\sqrt{3}\; P_x 
\beta_2\; sin\theta +  {  P_z  \over \sqrt{2}} \;\beta_3, \\
\label{t11}
C\,t^1_{\pm 1}&=& 
\;{9  \,i\,\alpha_3\; sin2\theta \over 2\sqrt{10}}
 \pm  \, {P_x \pm i P_y \over 2} \sqrt{3} \,\beta_4 \;cos\theta,\\
 \label{t20}
C\,t^2_0 &=& 
 \;\frac{(\alpha_4 - 9 \; \alpha_5 \, cos^2\theta)}{\sqrt{6}} +\,P_y
\,\beta_1\, sin\theta, \\
\label{t21}
C\,t^2_{\pm 1} &=&  
 \;\pm{3\alpha_6 \; sin2\theta \over 2} - i\,{P_x \pm i P_y \over \sqrt{2}} \sqrt{3}\,\beta_5\, cos\theta, \\
 \label{t22}
C\,t^2_{\pm 2} &=&  
 \;-{3\alpha_7 \; sin^2\theta \over 2}  \pm i\,{P_x \pm i P_y \over \sqrt{2}}\,
\sqrt{3} \, \beta_1 \; sin\theta,
\end{eqnarray}
where the factor $C$ is given by
\begin{equation}
C = 64\sqrt{3} \, \pi^2{ d\sigma \over dWd\Omega }
\end{equation} 
and $P_x, P_y, P_z$ denote components of the beam polarization ${\boldsymbol P}$.
The co-efficients $\alpha_3 , .. ,\alpha_7$  and $\beta_2,...,\beta_5$ are given by
\begin{eqnarray}
\alpha_3 &=&\Im( f_2f_3^*), \\
\alpha_4 &=&  |f_1|^2 + 3|f'|^2+ {9 \over 10} [|f_5|^2 +
{3 \over 5 }|f_6|^2 -  2\sqrt{10} \Re( f_4 f_5^*) \nonumber \\ &-&  6 \sqrt{2} \Re( f_4 f_6^*) -
{6 \over \sqrt{5} }\Re( f_5f_6^*)], \\
\alpha_5 &=& |f_2|^2+{3 \over 10}|f_3|^2-{2 \over \sqrt{10}}\Re( f_2 f_3^*),\\
\alpha_6 &=& |f_2|^2-{1 \over 5}|f_3|^2-{1\over \sqrt{10}}\Re( f_2 f_3^*)\;;\;
\alpha_7 = |f'|^2, \\
\label{beta2}
\beta_2 &=& \Re( f_1 f'^*) \;\;;\;\;
\beta_3 = |f_1|^2, \\
\beta_4 &=& \Re( f_1 f''^*)\;\;;\;\;
\beta_5 = \Im( f_1 f''^*). 
\end{eqnarray}

It is worth noting 
that the $\alpha_k$ and $\beta_k$ are bilinears in the partial wave amplitudes $f_k$ or 
equivalently the $R_k$, as explicited  through the Eq.\eqref{eq:fjb}.  They may also
be expressed, if necessary,  as bilinears in terms of the 
$M_k$ using Eq.\eqref{eq:fjb} and  then using Eq.\eqref{eq:gma} to express $R_k$ in terms of $M_k$.
For ready identification the notation $\alpha_k$ is employed to denote the bilinears which govern
measurements with unpolarized beam, while $\beta_k$ denote bilinears which govern observables 
in   polarized beam experiments.

It may be noted that $f_1, f_2, f_3$ lead to a singlet spin state of the two nucleons
in the final state, whereas $f_4,f_5,f_6$ lead to a triplet state. As such the two sets do not mix,
when no observations are made with regard to the spins of the two nucleons in the 
final state. Moreover, since $f_4, f_5$ and $f_6$ lead to the production of 
$s-$wave meson, their presence contributes only to the isotropic terms in the 
unpolarized differential cross-section given by Eq.\eqref{updcs} and tensor polarization
$t^2_0$ given by Eq.\eqref{t20}. As such it is difficult to estimate empirically the partial 
wave amplitudes $f_4, f_5, f_6$ individually along with $f_1, f_2, f_3$.

 If the    contributions of $f_4, f_5, f_6$ are neglected,  we have
 \begin{equation}
\alpha_0= \alpha_4 = |f_1|^2 + 3|f'|^2.
\end{equation}

Clearly,  $\alpha_1, \, \alpha_0$ and $\alpha_2$ can be determined from the experimental study of the angular
distribution of the unpolarized differential cross-section given by Eq.\eqref{updcs}. 
Moreover, if we set ${\boldsymbol P} = 0$, in Eqs. \eqref{t10} - \eqref{t22}, all the terms with coefficients 
$\beta_k, \; k = 1, \ldots, 5$ reduce to zero,   $C$ reduces to 
\begin{equation}
C_0 = 64\sqrt{3} \, \pi^2  { d\sigma_0 \over dWd\Omega }
\end{equation} 
and $t^k_q$ with ${\boldsymbol P} = 0$ may be denoted by  as $(t^k_q)_0$. 
The study of angular distribution of $C_0\,(t^2_0)_0$ with unpolarized beam determines $\alpha_4$
and $\alpha_5$.        
The $\alpha_3, \alpha_6$ and $\alpha_7$ are determinable from measurements 
of  $C_0\,(t^1_{\pm 1})_0, C_0\,(t^2_{\pm 1})_0$ and  $C_0\,(t^2_{\pm 2})_0$ respectively. 
 Thus,  one can  estimate empirically
\begin{eqnarray}
|f_1|^2 &=& \alpha_0+{3 \over 5} \alpha_2 - 3\, \alpha_5 \\
|f_2|^2 &=& {1 \over 90}[45 \,\alpha_5 + 36\, \alpha_6 -7\alpha_2]\\
|f_3|^2 &=& {1 \over 9}[ 20 \,(\alpha_5 -  \alpha_6)-\alpha_2 ],
\end{eqnarray}
from the measurements employing  an unpolarized beam.

It is seen from Eq. \eqref{ap} that the measurement of analyzing power $A_y $ in the 
Madison frame \cite{sat}  determines 
\begin{equation}
\Im(f_1^*f') = \beta_1
\end{equation}
when the beam is polarized transverse to the reaction plane i.e., along the y-axis.

The Cartesian components of beam  to meson spin transfers 
may be defined following \cite{ohel} through
\begin{eqnarray}
\frac{d\sigma}{dWd\Omega}P^\omega_i &=& \frac{d\sigma_0}{dWd\Omega} 
\sum_{j=x,y,z} \left((P^\omega_{i})_0+ K^{i}_j P_j\right), \\
\frac{d\sigma}{dWd\Omega}P^\omega_{ij} &=& \frac{d\sigma_0}{dWd\Omega} 
\sum_{k=x,y,z} \left((P^\omega_{ij})_0+ K^{ij}_k P_k\right),
\end{eqnarray}
where the Cartesian components $P^\omega_i, \, i = x,y,z$ of the vector 
polarization and $P^\omega_{ij}, \, i,j = x,y,z$ of the tensor polarization of the  $\omega$ 
meson with spin-1 are given, following \cite{sat}, in terms of the Fano statistical tensors $t^1_q$ and 
$t^2_q$ respectively, while $(P^\omega_i)_0, (P^\omega_{ij})_0$ are given respectively by 
$(t^1_q)_0$ and $(t^2_q)_0$  leads to 
\begin{eqnarray}
\label{kxxy}
C_0\,K^{xx}_y &=&  -2\,\sqrt{2}\,\beta_1\,sin\theta,\\
\label{kyyy}
C_0\,K^{yy}_y &=& \sqrt{2}\,\beta_1\,sin\theta,\\
\label{kzzy}
C_0\, K^{zz}_y &=& \sqrt{2}\,\beta_1\,sin\theta,
\end{eqnarray}
which add up to zero and 
\begin{eqnarray}
\label{kxyx}
C_0\,K^{xz}_y  &=&  - C_0\, K^{yz}_x  =  -\frac{3}{\sqrt{2}}\, \beta_5\,cos\theta, \\
\label{kxzy}
C_0\,K^{xy}_x  &=&  \frac{3}{\sqrt{2}}\, \beta_1\,sin\theta, 
\end{eqnarray}
all other $K^{ij}_k$ being zero. The non-zero $K^i_j$ are given by
\begin{eqnarray}
\label{kxx}
C_0\, K^x_x &=& \,C_0\,K^y_y = -\beta_4 cos\theta, \\
\label{kzx}
C_0\, K^z_x &=& \sqrt{2}\,\beta_2 \,sin\theta,\\
\label{kzz}
C_0\, K^z_z &=& \frac{1}{\sqrt{3}}\,\beta_3.
\end{eqnarray}

It is thus seen that $\beta_1$ can be determined not only from  the analyzing power $A_y$
given by Eq. \eqref{ap} but also from the tensor  polarization  of the $\omega$ given by Eqs. 
\eqref{t20} and \eqref{t22} or  equivalently from Eq.s \eqref{kxxy}-\eqref{kzzy}  and \eqref{kxzy} 
when the beam is polarized.  
One can determine $\beta_5$ from Eq. \eqref{t21} or equivalently from Eq. \eqref{kxyx}. The empirical estimates of the 
bilinears $\beta_2, \beta_3, \beta_4$ are obtainable from Eq.s \eqref{kzx},  \, \eqref{kzz}, \,\eqref{kxx} respectively 
or equivalently from Eq.s \eqref{t10} and \eqref{t11}.

Without any loss of generality one may assume $f_1$ to be real and express
$ f_2=|f_2|\,exp\,(i\varphi_2), \; f_3=|f_3|\,exp\,(i\varphi_3)$ so that one can determine 
\begin{eqnarray}
\label{sin2}
sin\varphi_2 = -{ 2\beta_1 + \beta_5 \over 3 |f_1||f_2|} \; &;& \; 
cos\varphi_2 = { 2\beta_2 + \beta_4 \over 3 |f_1||f_2|} \\
\label{sin3}
sin\varphi_3 = { \beta_5 - \beta_1 \over 3 |f_1||f_3|} \; &;& \; 
cos\varphi_3 = { \beta_2 - \beta_4 \over 3 |f_1||f_3|}. 
\end{eqnarray}

Thus it is possible to empirically determine $f_1, f_2, f_3$ along with their 
relative phases, without  any discrete ambiguities.

It may perhaps be pointed out also that $|f_1|^2,\, |f_2|^2,\, |f_3|^2, \, |f'|^2 $ and $ |f''|^2 $ 
as well as the relative phase between $f_2$ and $f_3$ can be determined using 
\begin{eqnarray}
\Re( f_2 f_3^*) &=& {5 \over 9 \sqrt{10}}[ 2 \,(\,\alpha_5 +  \alpha_6)-\alpha_2 ],\\
\label{alp3}
\Im(f_2f_3^*) &=& \alpha_3,
\end{eqnarray}
from the measurements employing an unpolarized beam.  However, the determination 
of $\alpha_3$ in \eqref{alp3} involves measurement of vector polarization
of $\omega$.  The measurement of vector polarization cannot be
carried out using the dominant $3\pi$ decay mode of $\omega$ \cite{gr4}.
The determination of the relative phase of $f'$ with $f_1$ 
without any trigonometric ambiguity involves determination of $\beta_1$ and $\beta_2$
i.e., the measurement of the analyzing power given by \eqref{ap} 
or any of the the spin transfers given by
Eq.s \eqref{kxxy}, \eqref{kyyy}, \eqref{kzzy} and the spin transfers given by Eq. \eqref{kxzy}.
Likewise the determination of the relative phase between $f_1$ and $f''$
involve the determination of spin transfers  $K^x_x = K^y_y$ and hence $\beta_4$
in Eq. \eqref{kxx} and $K_y^{xz}=-K_x^{yz}$ and hence $\beta_5$ in Eq. \eqref{kxyx}.
It is also interesting to note that  $|f_1|^2 = \beta_3$ in Eq. \eqref{beta2} is directly determined by 
the spin transfer $K^z_z$ given by  Eq. \eqref{kzz}.
With $|f_1|^2$ thus known,  $|f'|^2$ and $|f''|^2$ can be determined directly from
the measurements of the unpolarized differential cross-section at $\theta = \pi/2$
and $\theta = 0 $ or $ \pi$.  

It may also be pointed out that the measurement of the beam analyzing power and the tensor polarization of $\omega$
employing a polarized beam determine only $sin\varphi_2$ and $sin\varphi_3$, whereas,  
determination of $cos\varphi_2$ and $cos\varphi_3$ in Eq.s 
\eqref{sin2} and \eqref{sin3} necessarily involves measurements 
of the vector polarization of $\omega$ employing a polarized beam.

It was pointed out earlier \cite{gr3} that the decay mode $\omega \to \pi^0\,\gamma$
with the branching ratio of $8.9\%$ may be utilized to measure vector polariztion of 
$\omega$. It is encouraging to note that WASA \cite{was}  at COSY  is expected to facilitate the experimental study of 
$pp \to pp\omega$ via the detection of $\omega \to \pi^0 \gamma$ decay. It is to be  
noted, however,  that the determination of the vector polarization of $\omega$ involves measuring the 
circular polarization asymmetry of the radiation, whereas the angular distribution of 
the intensity of the radiation provides information on the tensor polarization.

Finally, we may note that the measurement \cite{abl} of $A_y$ compatible with zero 
does not necessarily imply  $f_2 = f_3 = 0$, but may indicate also that the 
relative phase between $f_1$ and $f'$ is zero. Since the already observed anisotropy
in the angular distribution of the   unpolarized differential cross-section 
invalidates the assumption that $f_2 = f_3 = 0$, it is very likely that the measurement \cite{abl}  at 
$\epsilon = 129 MeV$ indicates only that the relative phase of $f'$ with respect 
to $f_1$ is zero at that energy.

\begin{acknowledgments}
We thank Professor K T Brinkmann for encouraging us to look into this problem and for 
private communications from time to time with 
regard to the progress of the experiments at COSY .
One of us, (JB)  thanks Principal T. G. S. Moorthy and the Management of
K. S. Institute of technology  for encouragement.
\end{acknowledgments}

\end{document}